# DIFFUSE SCATTERING OF NEUTRONS IN A WAVE RESONATOR

E.D. Kolupaev[1], V.D. Zhaketov[1,2], Yu.V. Nikitenko[1]

[1]*Frank Laboratory of Neutron Physics, Joint Institute for Nuclear Research, 141980 Dubna, Moscow Region, Russia*

[2] *National Research University Moscow Institute of Physics and Technology, Moscow, Russia*

## Abstract

In fundamental experiments with neutrons, the neutron flux and the neutron storage time in the measuring setup are of primary importance. These quantities can be increased by using a storage device for neutrons generated by a pulsed source. In a storage device with material walls, both parameters are determined by the probabilities of neutron absorption and diffuse scattering upon reflection from the storage walls, as well as by the neutron decay probability. This work considers a neutron measurement method and presents the results of an experimental determination of the probability of diffuse neutron scattering in a wave resonator.

## INTRODUCTION

At present, neutron experiments aimed at determining the neutron lifetime with respect to beta decay [1, 2], the probabilities of neutron transformation into an antineutron [3–7], and into a mirror neutron [8–12] are of considerable relevance. A ring storage trap for cold neutrons can be used to determine these fundamental quantities [13–16]. In a storage trap with material walls, the probabilities of neutron absorption by atomic nuclei and diffuse neutron scattering by surface roughness limit both the neutron flux and the neutron storage lifetime.

The probability of diffuse neutron scattering exceeds the probability of neutron absorption in the walls of a storage trap by two to three orders of magnitude for neutron-reflecting materials such as copper and beryllium, for example. In this context, it is necessary to measure roughness parameters and to reduce the probability of diffuse neutron scattering. Roughness parameters can be determined by measuring X-ray reflection. However, X-ray absorption is comparable to diffuse scattering and is sufficiently large in comparison with neutron absorption, which reduces the accuracy of roughness-parameter determination. Therefore, for the problem of using a neutron storage trap, the application of a neutron-based measurement method is appropriate.

On the other hand, the probability of diffuse neutron scattering is two orders of magnitude lower than the probability of specular total reflection, which occurs when the neutron kinetic energy in the direction perpendicular to the surface is lower than its potential energy of interaction. Consequently, measuring the probability of diffuse scattering against the background of specular reflection is a practically challenging task. To determine the probability of diffuse neutron scattering, a measurement method based on a neutron wave resonator was proposed [17–18], which makes it possible to increase the sensitivity of the measurements. The method consists in using multiple reflection of the neutron wave from the interlayer boundaries of the resonator, whose layers are made of a material that can be used for the walls of a neutron storage trap.

This work considers the measurement method and presents the results of an experimental determination of the probability of diffuse neutron scattering in a resonant layered structure, namely a wave resonator.

## MEASUREMENT METHOD

The propagation of a neutron at small values of the wave vector $k$, in particular in a spatially ordered medium at $k < \pi/L$, where $L$ is the distance between scattering centers (atoms, atomic nuclei, clusters), is described by the complex interaction potential of neutrons with the medium,

$$U = V - iW = V(1 - i\eta)$$

[19]. The real $V$ and imaginary $W$ parts of the potential are expressed through the corresponding neutron wave vectors $k_V$ and $k_W$:

$$V = \frac{\hbar^2 k_V^2}{2m}, \qquad W = \frac{\hbar^2 k_W^2}{2m}, \qquad (1)$$

where

$$k_V^2 = 4\pi N b, \qquad k_W^2 = \frac{m}{\hbar} N\sigma v.$$

Here, $\hbar$ is Planck's constant, $m$ is the neutron mass, $b$ is the neutron scattering length for a scattering center, $\sigma$ is the cross section for neutron capture and scattering by a scattering center and for scattering by the medium, $N$ is the density of scattering centers, and $v$ is the neutron velocity.

A neutron wave resonator is a three-layer resonator structure (Fig. 1), in which the outer layers I and III have a higher real part of the neutron–medium interaction potential $V$ than the middle layer II, where neutron absorption is small. Neutrons undergo sub-barrier reflection from layers I and III, which occurs when the condition $k_0 < k_v$ is satisfied, where $k_0$ is the component of the neutron wave vector in vacuum perpendicular to the interfaces.

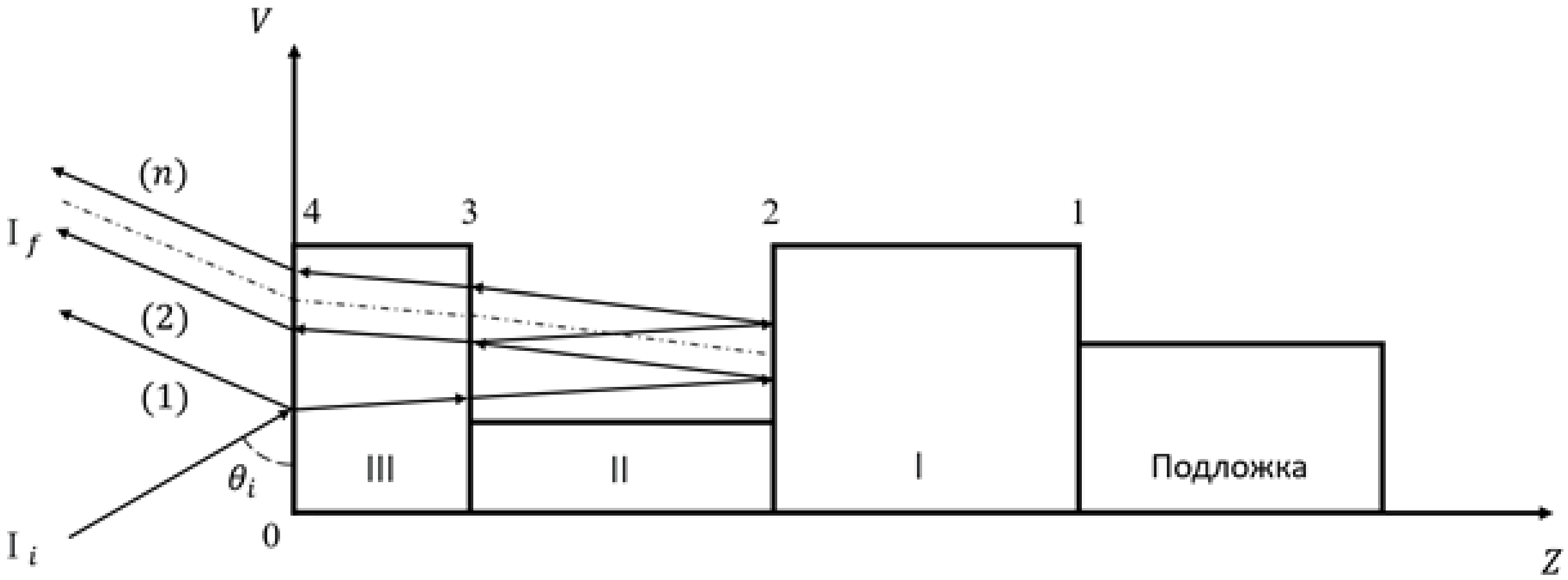

Fig. 1. Dependence of the real component of the potential $V$ on the coordinate $Z$, measured from the surface into the depth of the structure, for a three-layer structure. Numbers I–III denote the layers of the structure, and numbers 1–4 denote the interfaces between the layers. The orders of the reflected neutron waves are denoted by $(1)$–$(n)$.

Figure 2 shows the dependence of the reflection coefficient $R$ on $k_0$ for an ideal structure, i.e., a structure with rectangular layer potentials, Cu(300 Å)/Al(400 Å)/Cu(1000 Å)/glass(5 mm). On the plateau where $R = 1$, a dip is observed at the resonant wave-vector value $k_{\mathrm{res}} = 0.00686$ Å$^{-1}$, which is associated with multiple passage of neutrons through the intermediate layer II and their sub-barrier reflection from layers I and III.

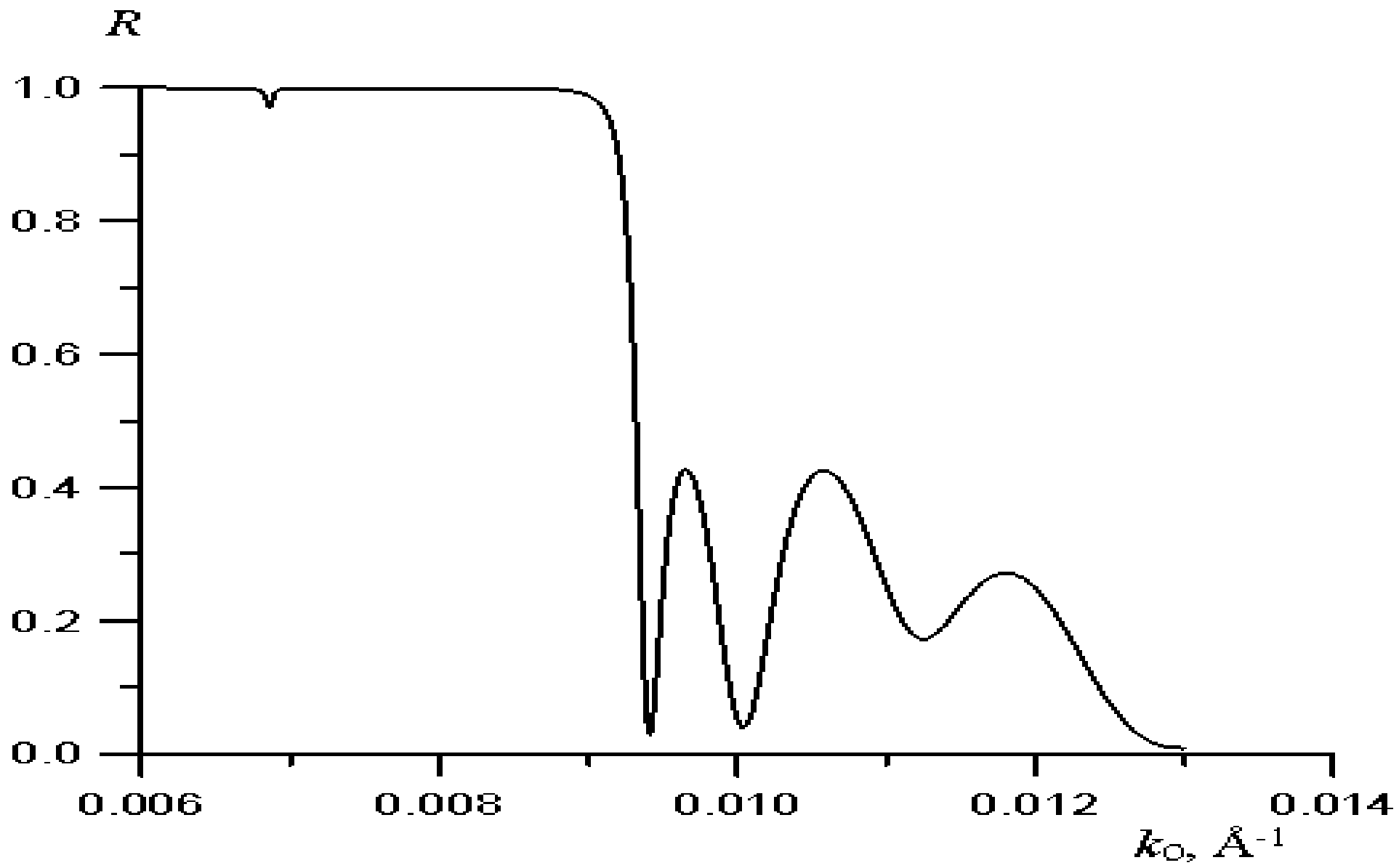

Fig. 2. Dependence of the reflection coefficient $R$ on the perpendicular component of the wave vector for the Cu(300 Å)/Al(400 Å)/Cu(1000 Å)/glass structure.

For the neutron absorption coefficient in the structure, the following relation holds [18]:

$$M = \int \left|\frac{\psi(k,z)}{\psi_0(k_0)}\right|^2 \frac{k_w^2(k,z)}{k_0}\, dz, \qquad [2]$$

where $\psi_0(k_0)$ is the wave function of the neutron incident on the structure, $\psi(k,z)$ is the neutron wave function in the structure, and $k$ is the perpendicular component of the wave vector in the structure.

As follows from Eq. (2), the value of $M$ is determined by the neutron density $n(z) = |\psi(z)|^2$, or, more precisely, by the flux $j(z) = vn(z)$, since the velocity $v$ enters the expression for $k_w^2$. In the resonator structure, $n(z)$ increases at resonant values of the wave vector, thereby increasing the coefficient $M$.

For sub-barrier reflection, when $W \ll V$ and $|\psi_0|^2 = 1$, Eq. (2) is transformed into the form [19]

$$M = \frac{2\eta k_0}{\left(k_v^2 - k_0^2\right)^{1/2}}. \qquad (2)$$

It follows from Eq. (3) that $M \to 0$ as $k_0 \to 0$, and $M = \eta$ when $k_0^2 = \frac{k_v^2}{5}$.

Figure 3 shows the coordinate dependences of the neutron density for different wave-vector values for the structures Cu(300 Å)/Al(400 Å)/Cu(1000 Å)/glass and Cu(300 Å)/Al(400 Å)/Be(1000 Å)/glass.

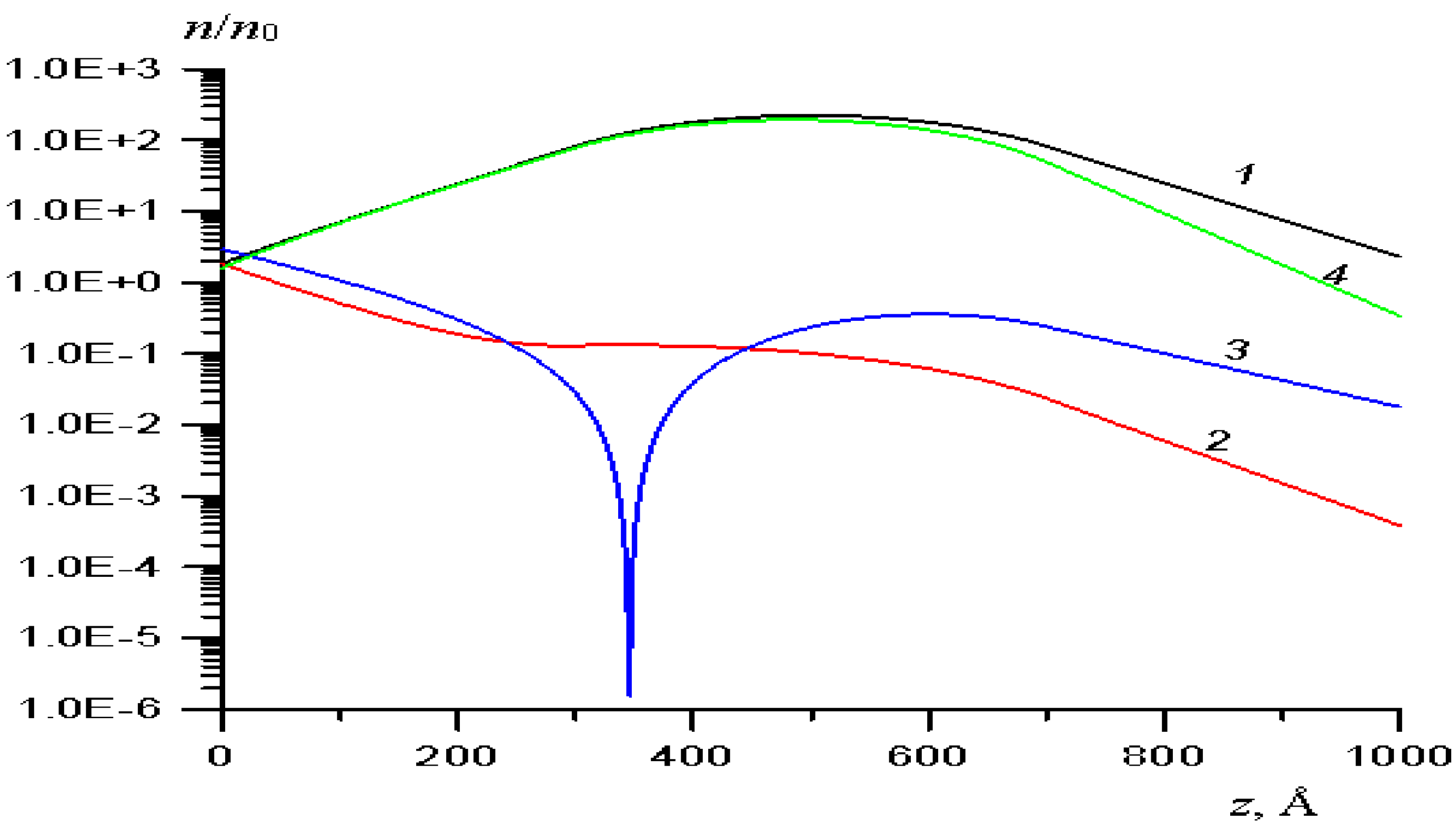

Fig. 3. Spatial dependence of $n(z)/n_0$, where $z$ is measured from the surface, for the Cu(300 Å)/Al(400 Å)/Cu(1000 Å)/glass structure at $k_0 = k_{\text{res}} = 0.00686$ Å$^{-1}$ (1), $k_0 = 0.006$ Å$^{-1} < k_{\text{res}}$ (2), and $k_0 = 0.008$ Å$^{-1} > k_{\text{res}}$ (3), and for the Cu(300 Å)/Al(400 Å)/Be(1000 Å)/glass structure at $k_0 = k_{\text{res}} = 0.00686$ Å$^{-1}$ (4).

At the resonant wave-vector value $k_0 = k_{\text{res}} = 0.00686$ Å$^{-1}$, curve 1, the density increases starting from the surface of layer III, reaches a maximum in the intermediate layer II, $300$ Å $< z <$ $700$ Å, and decreases with increasing $z$ in layer I, $z > 700$ Å. At $k_0 = 0.006$ Å$^{-1}$ $< k_{\text{res}}$, curve 2, the density decreases with increasing $z$. At $k_0 = 0.008$ Å$^{-1}$ $> k_{\text{res}}$, curve 3, a minimum is observed at $z = 340$ Å, in layer II, due to interference of counter-propagating waves.

At the resonant value of the wave vector, the density at the boundaries of the middle layer for the Cu(300 Å)/Al(400 Å)/Cu(1000 Å)/glass structure differs by 3% and is $\frac{n(z=300\text{ Å})}{n_0} = 80.962$ and $n(z = 700\text{ Å})/n_0 = 83.49$. As will be shown below, the practically equal density values lead to the extraction of averaged roughness parameters for the two boundaries of the middle layer.

In the Cu(300 Å)/Al(400 Å)/Be(1000 Å)/glass structure, the density is different: $n(z = 300\text{ Å})/n_0 = 77.36$, and $n(z = 700\text{ Å})/n_0 = 49.85$.

The difference in density values for this structure is 55% and is associated with different phases of the neutron reflection amplitude from the copper layer and the beryllium layer, which have different $k_v$ values. In this context, in the structure Cu(300 Å)/Al(400 Å)/Be(1000 Å)/glass, in which layers I and III are made of different materials—namely, layer I is made of beryllium and layer III of copper—the values of the roughness parameters of interfaces 2 and 3 can be determined independently.

To qualitatively describe the physical picture of neutron density, or flux, formation in the resonator, we present relations for the absorption coefficients in individual layers of the structure. For the neutron wave function in the first reflecting layer of the structure, the following relation holds:

$$\psi_1 = \frac{\psi_0 t_{32} t_{i1} \exp(ik_1)}{1 - r_{23} r_{i1}}, \qquad (4)$$

where

$$t_{32} = \frac{t_3 t_2}{1 - r_2 r_3}$$

is the transmission amplitude of the bilayer consisting of the third and second layers,

$$r_{23} = r_2 + \frac{t_2^2 r_3}{1 - r_2 r_3}$$

is the reflection amplitude from the bilayer consisting of the second and third layers,

$$r_{2(3)} = \frac{r_{i2(i3)}\left[1 - \left(1 - r_{i2(i3)}^2\right)\exp\left(2ik_{2(3)}d_{2(3)}\right)\right]}{1 - r_{i2(i3)}^2 \exp\left(2ik_{2(3)}d_{2(3)}\right)}$$

is the reflection amplitude from the second, or third, layer,

$$t_{2(3)} = \frac{\left(1 - r_{i2(i3)}^2\right)\exp\left(ik_{2(3)}d_{2(3)}\right)}{1 - r_{i2(i3)}^2 \exp\left(2ik_{2(3)}d_{2(3)}\right)}$$

is the neutron transmission amplitude through the second, or third, layer,

$$t_{i1} = \frac{2k_0}{k_0 + k_1}$$

is the transmission amplitude through the "vacuum–first layer" boundary,

$$r_{i1} = \frac{k_0 - k_1}{k_0 + k_1}$$

is the reflection amplitude from the "vacuum–first layer" boundary, $d_{2(3)}$ is the thickness of the second, or third, layer, and $k_{2(3)}$ is the component of the wave vector perpendicular to the interfaces in the second, or third, layer.

After transformations, at the resonant value of the wave vector $k_{\mathrm{res}}$, satisfying the relation $2k_{\mathrm{res}}d_2 + \varphi_1 + \varphi_3 = 2\pi$, where $\varphi_1$ and $\varphi_3$ are the phases of the reflection amplitudes from the first and third layers, respectively, we obtain the following expression for the neutron absorption coefficient in the first layer in the approximation $r_2 \approx 0$, $t_2 \approx \exp(ik_2 d_2)$, $r_3 \approx r_{i3} = |r_{i3}|\exp(i\varphi_3)$, and $t_{32} \approx t_3 t_2$:

$$M_1 = \chi_1 \mu_1 = \frac{|t_3|^2 \mu_1}{(\mu_1/2 + \mu_3/2 + \mu_2)^2}. \quad (5)$$

Here,

$$\chi_1 = \frac{|t_3|^2}{(\mu_1/2 + \mu_3/2 + \mu_2)^2}, \text{ and } \mu_{1,3} = \left(1 - \left|r_{i1,i3}\right|^2\right)$$

is the absorption coefficient in the first, reflecting [20], and third, enhancing [20], layers, respectively, $\mu_2 \approx 4d_2k_{2i}$ is the volume absorption coefficient in the second, phase-shifting [20], layer, and $k_2 = k_{2r} + ik_{2i}$. The enhancement factor of the absorption coefficient in the first layer, $\chi_1 = M_1/\mu_1$, is inversely proportional to the absorption coefficients in the layers $\mu_1$, $\mu_2$, and $\mu_3$.

For the third layer, in the same approximation, we have

$$M_3 \approx \chi_1\big(1 - \exp(-k_{i3}d_3)\big)\mu_3. \tag{6}$$

In Eq. (6), compared with Eq. (5), an additional factor appears, $1 - \exp(-k_{i3}d_3)$, which depends on the thickness $d_3$ of the third layer. Recall that the first layer is assumed to have infinite thickness, i.e., it is treated as a medium. With increasing $d_3$, the value of $M_3$ approaches $M_1$.

Figure 4 shows the dependences $M(z)$ for layers I, II, and III. It can be seen that at resonance $M_I \approx M_{III} > M_{II}$. the ratio of the absorption coefficients is approximately determined by the ratio of the parameters $\eta = (k_w/k_v)^2$, namely $\eta_I = \eta_{III} = 100\eta_{II}$.

At the same time, the attainable neutron density is inversely proportional to $\eta$. Therefore, aluminum, for which $\eta$ is an order of magnitude smaller than for copper, is used below as the middle layer. The aluminum layer is in the over-barrier neutron-reflection regime, whereas the copper and beryllium layers are in the sub-barrier regime.

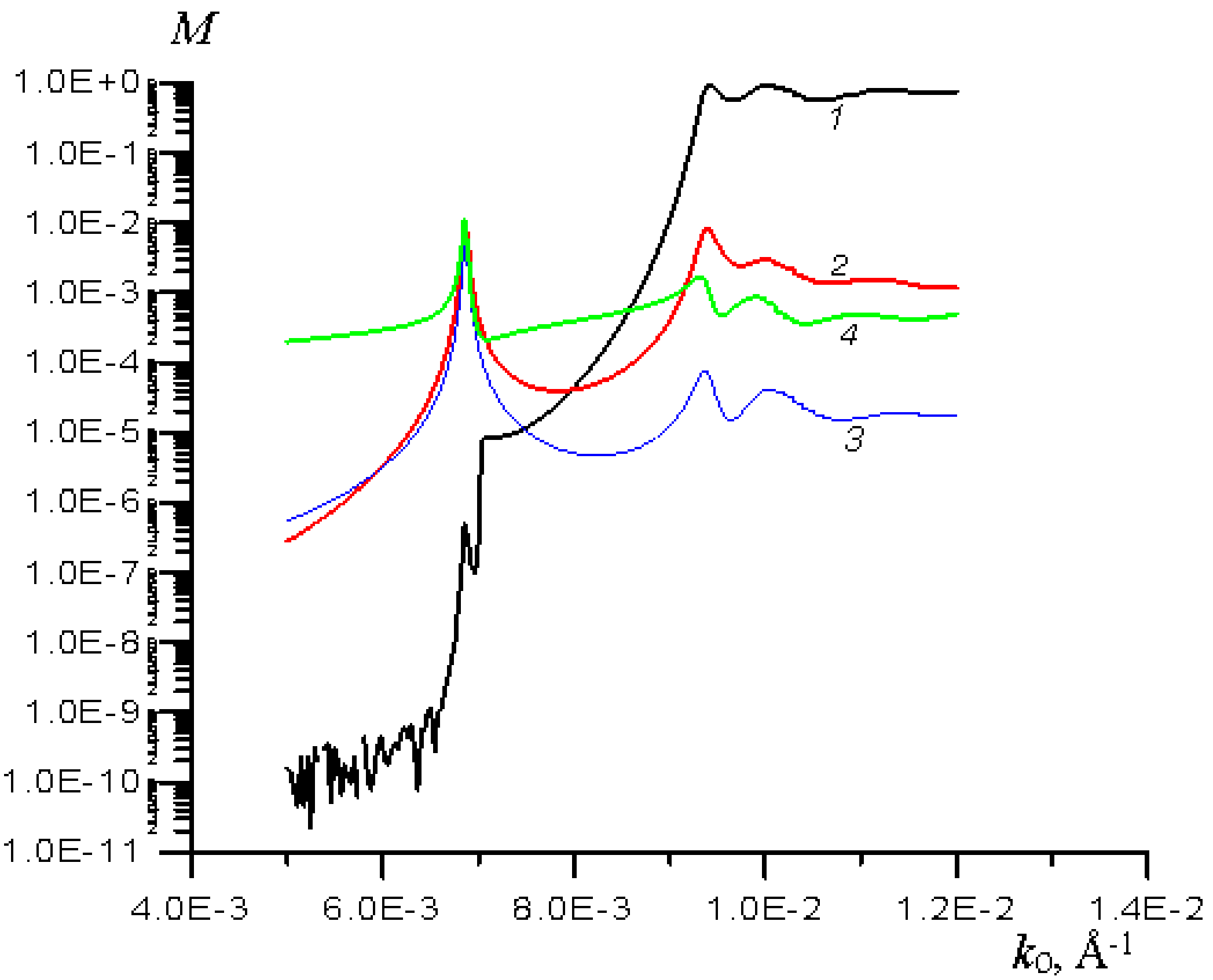

Fig. 4. Dependences of the absorption probability $M(z)$ in the substrate (1) and in layers I (2), II (3), and III (4).

## CALCULATION OF NEUTRON REFLECTION AND SCATTERING COEFFICIENTS

We now consider the probability of diffuse scattering. For the probability of diffuse scattering in the absence of roughness correlation at the interfaces along the depth direction of the structure, we have [21–23]

$$\frac{dS}{d\theta_f} = \left(\frac{1}{4\pi}\right)^2 \sum_{j=1}^{n} F_j \left|k_j^2 - k_{j+1}^2\right|^2 \left|\psi_{ji}(k_i)\right|^2 \left|\psi_{jf}(k_f)\right|^2, \quad (7)$$

where

$$F_j = \frac{2\pi\sigma_{jz}^2 L_{jx}^2}{\left(1 + q_x^2 L_{jx}^2\right)^{3/2}},$$

$\psi_{ji}(k_i)$ and $\psi_{jf}(k_f)$ are the neutron wave functions at the perpendicular components of the wave vector before and after scattering, $j$ is the interface number,

$$q_x = \frac{2\pi}{\lambda}\left(\cos\theta_i - \cos\theta_f\right),$$

$\theta_i$ is the grazing angle of the neutron incident on the structure, where the indices "i" and "0" are identical, $\theta_f$ is the grazing angle of the scattered neutron,

$$k_j^2 = 4\pi N_j b_j,$$

and $L_{jx}$ and $\sigma_{jz}$ are the correlation length and root-mean-square roughness amplitude at the interface.

It follows from Eq. (7) that the probability is the sum of scattering probabilities from roughness at individual interfaces. In the case of correlated roughness, however, the scattering amplitude is the sum of the amplitudes from individual interfaces, and for the probability we have [24]

$$\frac{dS}{d\theta_f} = \left(\frac{1}{4\pi}\right)^2 \left|\sum_{j=1}^{n} \psi_{jf}\,(k_{iz}) F_j^{1/2}\left(k_j^2 - k_{j+1}^2\right)\psi_{ji}\left(k_{fz}\right)\right|^2. \quad (8)$$

In the presence of neutron scattering and absorption in the structure, the flux-balance condition is

$$R + T + M + S = 1, \quad (9)$$

where $R$ is the neutron reflection coefficient from the structure, $T$ is the neutron transmission coefficient through the structure, $M$ is the neutron absorption coefficient in the structure, and $S$ is the neutron scattering coefficient by the structure.

Under sub-barrier reflection, when $T = 0$, the reflection coefficient is

$$R = 1 - M - S. \quad (10)$$

Let us introduce three characteristic quantities: the root-mean-square variation of the neutron grazing angle in the incident beam, $\sigma_\theta$; the interval of grazing angles of neutrons in the incident beam, $\theta_{d1}$–$\theta_{d2}$; and the interval of neutron detection angles by the detector, $\theta_{r1}$–$\theta_{r2}$. In the general case,

$$\sigma_\theta < (\Delta_r = \theta_{r2} - \theta_{r1}) < (\Delta_d = \theta_{d2} - \theta_{d1}),$$

and for the average reflection coefficient $R_a$ within the detection-angle range we have

$$R_a = \left[\int R(\theta) - \iint \frac{dS(\theta_i,\theta_{f1})}{d\theta_{f1}} d\theta_{f1}\right] f(\theta_i,\theta_a) d\theta_i + \iint \frac{dS(\theta_{f2},\theta_i)}{d\theta_i} d\theta_i f(\theta_{f2},\theta_a) d\theta_{f2}. \quad (11)$$

Here, $\theta_a$ is the mean value of the detection angle, and

$$f(\theta,\theta_a) = \exp\left[-\frac{(\theta-\theta_a)^2}{2\sigma_\theta^2}\right]$$

is the spectrometer resolution function. Integration over $\theta_i$ is performed within the detection-angle interval $\theta_{r1}$–$\theta_{r2}$. Integration over $\theta_{f1}$ is performed over the intervals $\theta_{r2}$–$\pi/2$ and 0–$\theta_{r1}$. Integration over $\theta_{f2}$ is performed over the intervals $\theta_{r2}$–$\theta_{d2}$ and $\theta_{d1}$–$\theta_{r1}$. In the experiment, the relations

$$\theta_{r1} = \theta_{d1}, \qquad \theta_{r2} = \theta_{d2}$$

were satisfied, so that the third term in the expression for $R_a$ is absent.

The coefficient $R_a$ can also be calculated recursively [25], knowing the reflection and transmission amplitudes of individual layers and of effective scattering regions of thickness

$$\Delta z = 2\sigma_z$$

at the layer interfaces [18].

## EXPERIMENTAL RESULTS AND DISCUSSION

The resonator structure Cu(300 Å)/Al(400 Å)/Cu(1000 Å)/glass, fabricated at PNPI, Gatchina, was selected for the study. The structure parameters were preliminarily characterized by measuring X-ray reflection. X-ray studies were performed using an Empyrean platform manufactured by Malvern Panalytical. Its parameters were as follows: wavelength 1.79 Å, grazing-angle interval 0.0526–2.15°, angular step $\Delta\theta = 0.001°$, source slit diaphragm 0.13 mm, source-to-sample distance 240 mm, sample-to-detector distance 240 mm, detector slit diaphragm 0.1 mm. The measurement geometry was $\theta - 2\theta$. The data were processed using the X'Pert Reflectivity software package. Fitting was performed using a segmented algorithm [26].

Figure 5 shows the results of approximating the experimental X-ray reflection data from the resonator structure. It can be seen that up to $\theta = 1.25°$, the calculated dependence agrees with the experimental data. At $\theta > 1.25°$, the calculated dependence is shifted relative to the experimental one in the angle $\theta$, which may be due to an error in determining the angle $\theta$.

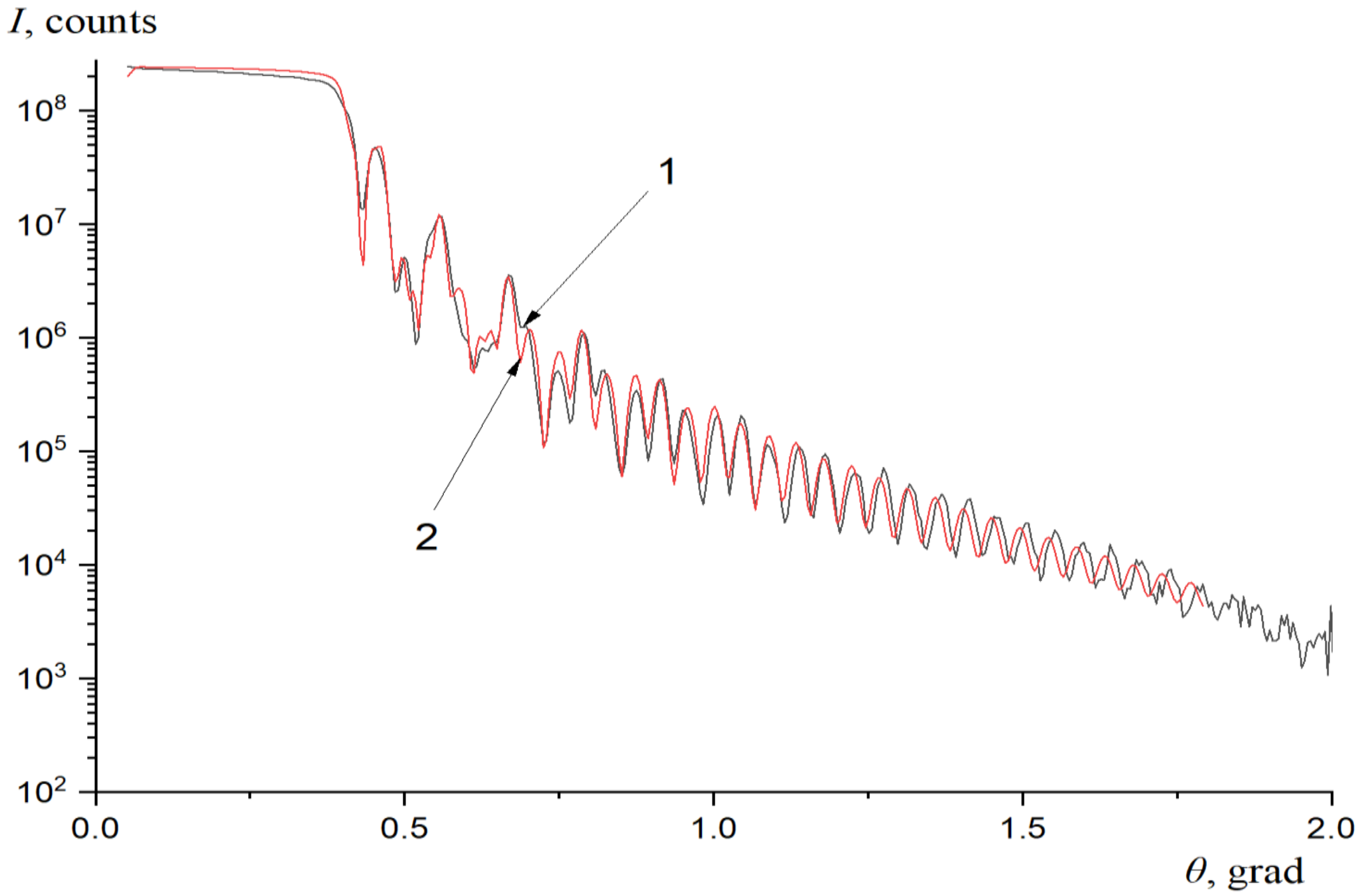

Fig. 5. Dependence of the experimental (1) and calculated (2) intensity of X radiation reflected from the Cu(30 nm)/Al(40 nm)/Cu(100 nm)/glass structure on the grazing angle of the incident beam.

Table 1 presents the structure parameters obtained by fitting the calculation to the experimental data. The layer thicknesses are close to their nominal values.

**Table 1**. Parameters of the Cu(30 nm)/Al(40 nm)/Cu(100 nm)/glass structure obtained by fitting the calculation to the experimental X-ray data.

| Layer No. | Material | Layer density, g/cm$^3$ | Layer thickness, Å |
|---|---|---|---|
| Substrate | Glass $Na_2O{\cdot}CaO{\cdot}6SiO_2$ | 2.52 | 5 mm |
| I | Cu | 8.94 | 1040 |
| II | Al | 2.42 | 401 |
| III | Cu | 8.91 | 338 |
| — | Vacuum | 0 | — |

Figure 6 shows the scheme of the neutron reflectometry measurements performed in the neutron wavelength range 1–15 Å. The neutron beam incident on the structure under study was collimated by diaphragms D1 and D2. The aperture width of D1 was $h_1 = 1.2$ mm. The role of diaphragm D2 was played by the sample itself. The diaphragm aperture width was determined from the relation $h_2 = L\sin(\theta)$, where $L = 85$ mm is the sample length and $\theta = 0.0035$ is the mean value of the neutron-beam grazing angle; it was $h_2 = 0.3$ mm. When measuring the spatial resolution of the neutron detector, the beam was collimated by diaphragm D3 with $h_3 = 0.1$ mm.

The distance from diaphragm D1 to the sample was 4500 mm, and the distance from the sample to the detector was 5030 mm. The spatial resolution of the detector was 1.56 mm.

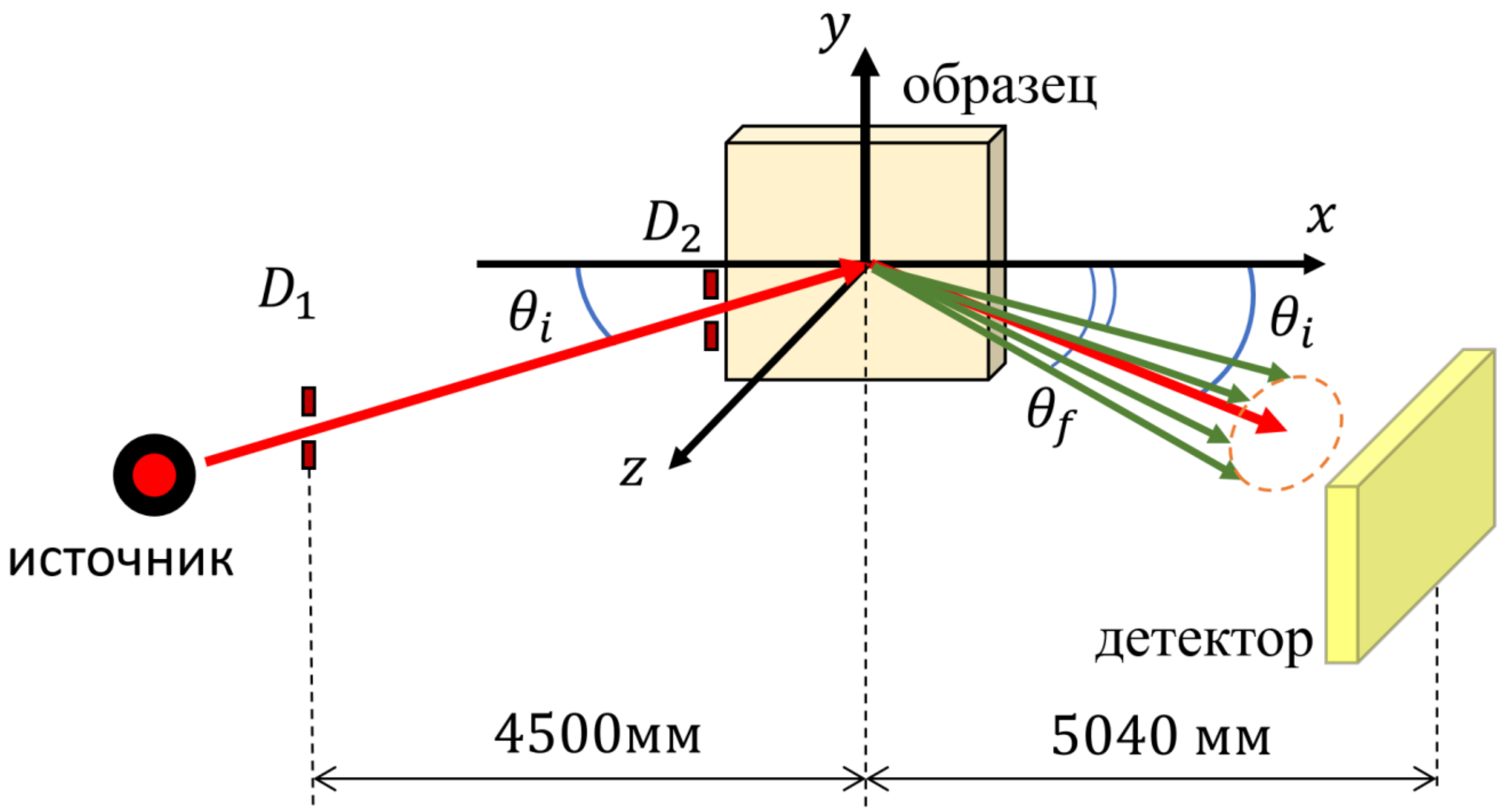

Fig. 6. Measurement scheme: D1,2 are diaphragms; $\theta_i$ and $\theta_f$ are the grazing angles of the neutron beams incident on and reflected from the structure.

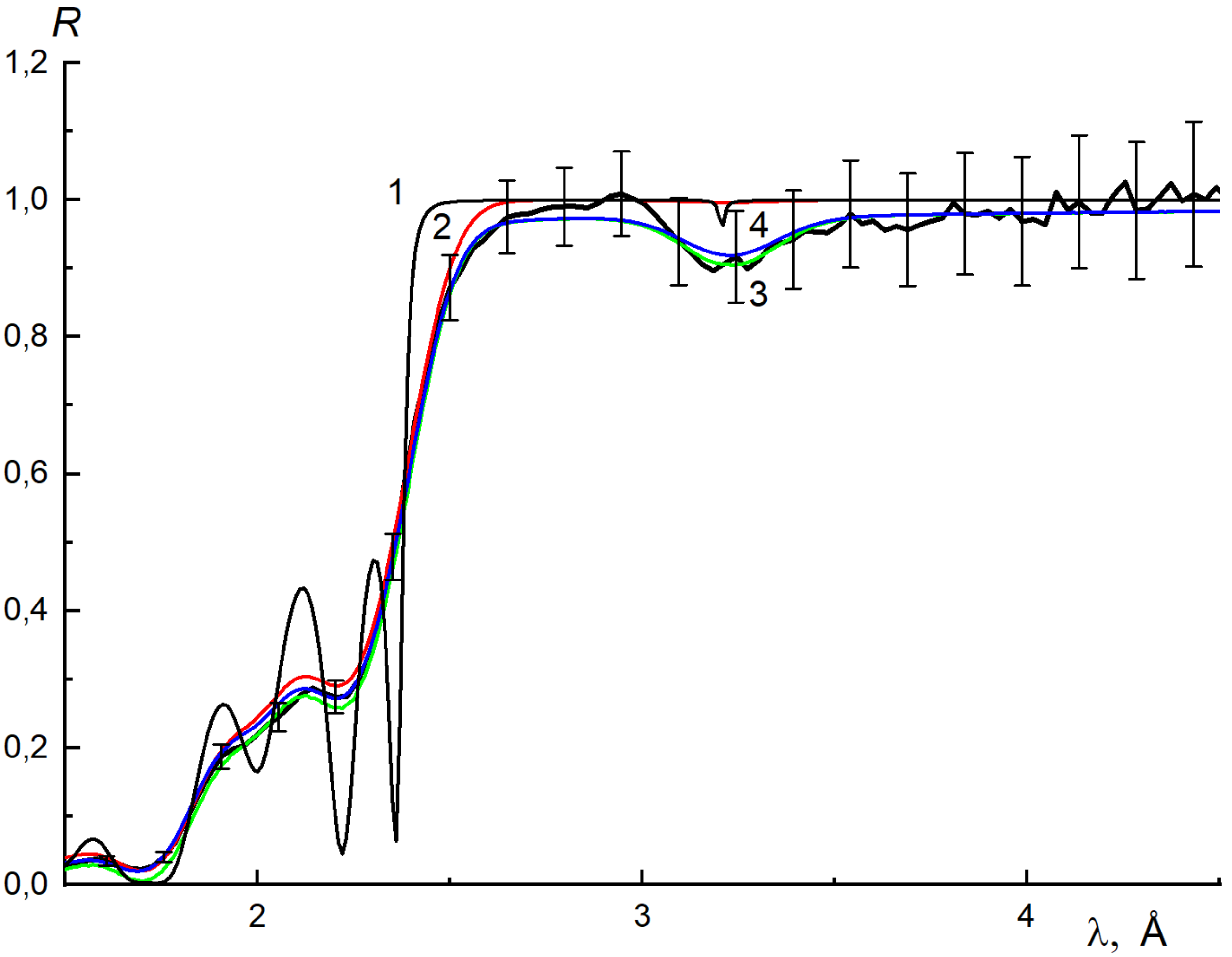

Fig. 7. Experimental data and calculated dependences (1–4) of the reflection coefficient on neutron wavelength at a grazing angle of 3.5 mrad and angular resolution $\sigma_\theta = 0.13$ mrad: (1) without taking into account resolution and roughness; (2) taking into account resolution and a detection interval of 2.84–4.16 mrad; (3) additionally to conditions (2), taking into account scattering by roughness using relation (11); (4) additionally to conditions (2), taking into account scattering by roughness using recursive relations and the values $k_{d,w}^2$ for boundaries 1–4.

Figure 7 presents the experimental data and calculated dependences (1–4) of the reflection coefficient on neutron wavelength at a grazing angle of 3.5 mrad and angular resolution $\sigma_\theta = 0.13$ mrad. In the calculations, the potential of the region of thickness $\Delta z = 2\sigma_z$ at the boundary between layers was represented as depending on the potentials of adjacent $j$ and $j+1$ layers. Thus, the real part of the potential was determined from the relation $V = \frac{V_j + V_{j+1}}{2}$. The imaginary part of the potential was represented as $W = W_{c,j} + W_{c,j+1} + W_d$, where $W_{c,j}$ and $W_{c,j+1}$ are the potentials due to the capture cross sections of the adjacent layers, and $W_d$ is the part of the imaginary potential due to diffuse scattering. The dip in Fig. 7 in the dependence of the reflection coefficient at $\lambda = 3.30$ Å is mainly associated with diffuse neutron scattering by roughness at the layer interfaces; the absorption coefficient here is 0.094. The absorption coefficient at the critical wavelength value of 2.88 Å is 0.023. Dependences 3 and 4 at resonance, at $\lambda = 3.30$ Å, differ by 1%. This agreement between the calculations performed using relation (10) and those performed using recursive relations confirms the possibility of representing the interface as a region in which the potential contains a term corresponding to diffuse scattering by roughness.

Table 2 presents the values of the parameters of the layers and interfaces. Boundaries 2–4 have identical roughness-parameter values. For boundary 1, we indicate the maximum parameter values that begin to worsen the fitting of the calculated dependence to the experimental one. In fact, the parameter values for boundary 1 may be smaller than or equal to those for boundaries 2–4. Assuming correlated interface roughness, the calculated parameters of boundaries 2–4 decrease by 1.8% to $\sigma_z = 11.8$ Å and $L_x = 4026$ Å.

Table 2. Neutron studies: values of the parameters of the layers and interfaces in the absence of interface correlation.

| Layer | Layer material | $k_u$, $nm^{-1}$ | $k_w$, $nm^{-1}$ | Layer thickness, Å | Interface | $\sigma_z$, Å | $L_x$, μm |
|---|---|---|---|---|---|---|---|
| Substrate | Glass: $Na_2O{\cdot}CaO{\cdot}6SiO_2$ | $7.0 \times 10^{-2}$ | $1.2 \times 10^{-3}$ | 5 mm | 1–Cu/glass | < 16 | < 0.55 |
| I | Cu | $9.05 \times 10^{-2}$ | $1.06 \times 10^{-3}$ | 1050 | 2–Al/Cu | $12 \pm 0.3$ | $0.41 \pm 0.01$ |
| II | Al | $5.1 \times 10^{-2}$ | $2.2 \times 10^{-3}$ | 401 | 3–Cu/Al | $12 \pm 0.3$ | $0.41 \pm 0.01$ |

| Layer | Layer material | $k_v$, nm$^{-1}$ | $k_v$, nm$^{-1}$ | Layer thickness, Å | Interface | $\sigma_z$, Å | $L_x$, μm |
|---|---|---|---|---|---|---|---|
| III | Cu | $9.05 \times 10^{-2}$ | $1.06 \times 10^{-3}$ | 338 | 4–vacuum/Cu | $12 \pm 0.3$ | $0.41 \pm 0.01$ |
| — | Vacuum | 0 | 0 | — | — | — | — |

In the resonator structure, diffuse scattering from individual interfaces is dominant in different regions of the full wave-vector range of neutrons incident on the structure. Indeed, for the first interface this is the range $k \gg (k_v(I), k_v(III))$. For the second and third interfaces, this is the vicinity of the resonant wave-vector value $k \approx k_{\text{res}}$. For the fourth interface, $k_v(II) \leq k \leq (k_v(I), k_v(III))$. Therefore, in principle, it appears possible to determine roughness parameters for all four interfaces between the layers if layers I and III are made with different $k_v$ values. For the experimentally studied structure, $k_v(I) = k_v(III)$. In this case, as noted in the explanation of the dependences in Fig. 3, only the averaged parameter values for boundaries 2 and 3 are determined. The roughness-parameter values for boundary 4 coincided with the averaged values for boundaries 2 and 3, which does not exclude roughness correlation for boundaries 2–4. For boundary 1, due to the large statistical error, we indicate only the maximum limiting values of the roughness parameters.

Thus, the loss coefficient due to diffuse scattering was

$$\mu_{\text{diff}} = 2.8 \times 10^{-3},\ 8 \times 10^{-4},\ \text{and}\ 4.1 \times 10^{-4}$$

for neutron wavelengths of 5, 10, and 20 Å, respectively. Alternatively, assuming that, as in the case of velocity-independent/wavelength-independent real and imaginary potentials, the relation

$$\mu = \frac{2\eta x}{(1 - x^2)^{1/2}},$$

where

$$x = \frac{\lambda_{\text{cr}}}{\lambda}, \qquad \lambda_{\text{cr}} = 2.36 \text{ Å},$$

is satisfied, one obtains for the corresponding wavelength values

$$\eta_{\text{diff}} = 1.44 \times 10^{-3},\ 1.64 \times 10^{-3},\ \text{and}\ 1.72 \times 10^{-3}.$$

The data obtained for $\eta_{\text{diff}}$ indicate that the imaginary potential responsible for diffuse scattering is not constant, but increases with increasing wavelength.

## CONCLUSION

The performed study shows that the loss coefficient at the vacuum–copper interface lies within the range $\mu_{\text{diff}} = 0.0004$–$0.03$ in the interval of $\lambda_{\text{cr}}/\lambda$ from 0.12 to 1, for roughness parameters $\sigma_z = 12$ Å and $L_x = 0.41\ \mu$m. This already makes it possible to carry out certain experiments in a neutron storage trap. For example, in determining the fundamental characteristics of the neutron, it is already sufficient to have values of $\mu = 10^{-3}$–$10^{-2}$. However, in order to increase the measurement capabilities and, consequently, to reduce $\mu_{\text{diff}}$, it is necessary either to decrease both roughness parameters, $\sigma_z$ and $L_x$, or to decrease $\sigma_z$ while increasing $L_x$ beyond the critical value at which the scattering probability begins to decrease. In the second case, diffuse scattering is concentrated near specular reflection, which corresponds to a decrease in the absorption coefficient within a given angular range due to an increase in the permissible number of neutron reflections during neutron propagation in a storage device. In this regard, reflection from a wall covered with a liquid surface layer, for which $L_x$ reaches a value of 1 mm, should be considered and investigated.

## REFERENCES

[1] Ezhov V.F. et al. // Pis'ma v ZhETF. 2018. Vol. 107. Iss. 11. P. 707.

[2] Serebrov A.P. // UFN. 2019. Vol. 189, No. 6. P. 635.

[3] Chetyrkin K.G., Kazarnovsky M.V., Kuzmin V.A., Shaposhnikov // Physics Letters. 1981. Vol. 99B. No. 4. P. 358.

[4] Lushchikov V.I., Popov A.B., Samosvat G.S., Taran Yu.V. // JINR Communications P3-81-313, Dubna, 1981.

[5] Baldo-Ceolin M. et al. // Physics Letters B. 1990. Vol. 236. No. 1. P. 95.

[6] Nesvizhevsky V.V., Gudkov V., Protasov K.V., Snow W.M., Voronin A.Yu. // EPJ Web of Conferences. 2018. Vol. 191. P. 01005.

[7] Mistead D. // arXiv: 2015.1510.015v1.

[8] Pokotilovsky Yu.N. // Phys. Lett. 2006. Vol. 639. P. 214–217.

[9] Okun L.B. // UFN. 2007. Vol. 177. No. 4. P. 397.

[10] Serebrov A.P. // Phys. Lett. 2008. Vol. 663. P. 181–185.

[11] Broussard L.J., Bailey K.M., Bailey W.B. et al. // EPJ Web of Conf. 2019. Vol. 219. P. 07002.

[12] Addazi A., Anderson K., Ansell S. et al. // J. Phys. G: Nucl. Part. Phys. 2021. Vol. 48. P. 070501.

[13] Kugler K.J., Moritz K., Paul W., Trinks U. // NIM A. 1985. Vol. 228. P. 240.

[14] Paul W., Anton F., Paul L., Paul S., Mampe W. // Z. Phys. C. 1989. Vol. 45. P. 25.

[15] Nikitenko Yu.V. Cold Neutron Storage Device. Patent No. 2772969, 30.05.2022.

[16] Nikitenko Yu.V. Storage Device for Cold and Very Cold Neutrons, P13-2023-28, Dubna, 2023.

[17] Nikitenko Yu.V. Method for Measuring the Probability of Neutron Absorption during Sub-Barrier Reflection from a Surface and a Structure for Its Implementation. Patent No. 2761053, 02.12.2021.

[18] Nikitenko Yu.V. Absorption and Scattering of Neutrons during Sub-Barrier Reflection from a Surface. Journal of Surface Investigation: X-Ray, Synchrotron and Neutron Techniques. Vol. 18. Supplement, 2024.

[19] Shapiro F.L. Neutron Investigations. Moscow: Nauka, 1976.

[20] Aksenov V.L., Nikitenko Yu.V. // Physica B. 2001. Vol. 297. P. 101–112.

[21] Sinha S.K., Sirota E.B., Garoff S., Stanley H.B. // Phys. Rev. B. 1988. Vol. 38. P. 2297.

[22] Deak L., Bottyan L., Nagy D.L., Spiering H., Khaidukov Y.N., Yoda Y. // Phys. Rev. B. 2007. Vol. 76. P. 224420.

[23] Khaydukov Yu., Morari R., Soltwedel O., Keller T., Christiani G., Logvenov G., Kupriyanov M., Sidorenko A., Keimer B. // Journal of Applied Physics. 2015. Vol. 118. P. 213905.

[24] Nikitenko Yu.V., Syromyatnikov V.G. Polarized Neutron Reflectometry. Moscow: Fizmatlit, 2013.

[25] Ignatovich V.K. Neutron Optics. Moscow: Fizmatlit, 2006.

[26] Atkinson A.C. A segmented algorithm for simulated annealing // Statistics and Computing. 1992. Vol. 2.